\documentclass[a4paper,11pt, hyphens]{article}
\usepackage{jinstpub} 
\usepackage{lineno}
\usepackage{float}

\title{\boldmath Characterization of micro-SPECT system based on Timepix detector}

\author[a]{V.~Rozhkov}
\author[a, b]{, I.~Hern\'{a}ndez}
\author[c]{, A.~Leyva}
\author[a]{, A.~Zhemchugov}

\affiliation[a]{Joint Institute for Nuclear Research, Dubna, Russia}
\affiliation[b]{Isotopes Center, Mayabeque, Cuba}
\affiliation[c]{Center of Technological Applications and Nuclear Development, Havana, Cuba}

\emailAdd{rozhkov@jinr.ru}

\abstract{In this work, the characteristics of a prototype SPECT system based on the Timepix readout chip, with a MURA type encoding mask, were evaluated. The set-up has a small field of view and can be used in preclinical studies of drugs in small laboratory animals. Despite many existing test protocols developed and described in pertinent documents of national standard bodies and IAEA recommendation, they are not suitable for microtomographic systems based on semiconductor pixel detectors due to different detector technology, high spatial resolution and small area of interest. To measure their characteristics, special phantoms were developed, with a small "hot~region". 
\par Such micro-SPECT parameters as spatial resolution, contrast, linearity and system efficiency were studied using $^{99m}Tc$ source. The detector calibration and data preprocessing are described.}


\begin{document}
\maketitle
\flushbottom

\section{Introduction}
\label{sec:intro}

\par In recent years, the use of radio-tracers has found wide application in diagnostic medicine and preclinical drug testing, where information about the dynamics of labeled molecule accumulation is obtained. In such molecules, some stable atoms are replaced by radioactive ones that are chemically equivalent. By binding these compounds to specific proteins, it is possible to deliver them to targeted organs. Depending on the isotope decay channel, either PET or SPECT technology can be employed to visualize drug distribution. The most common devices for SPECT imaging are Anger cameras \cite{anger}. However, their low spatial resolution is suboptimal for preclinical studies on laboratory animals. Compton cameras could offer a solution, though their applicability is constrained by photon energy requirements—typically around 200~keV or higher—thus limiting the selection of isotopes and detectors suitable for this approach. The photon emitters most widely used in SPECT, such as $^{99m}Tc$, $^{66}Ga$, and $^{123}I$, emit radiation in the 30 to 180 keV range. For this energy window, imaging tools with a small field of view are more appropriate. These are particularly well-suited for research involving small rodents. Their spatial resolution generally ranges from 0.2 mm to 2 mm and is strongly influenced by the collimator design. Pinhole collimators and their modifications are frequently applied \cite{e}. Multi-pinhole configurations, in particular, offer improved sensitivity while preserving high spatial resolution.

\par This article presents the characteristics of a SPECT imaging system based on a Timepix detector with a CdTe sensor, using MURA-type coding apertures collimators.

\section{The SPECT setup}

\par The SPECT setup studied in this paper is based on the use of a hybrid pixel detector made of a Timepix readout chip, with a 2 mm thick CdTe sensor. A MURA coding aperture with 31~rank basic pattern was used as a collimator \cite{i}.

\par The Timepix chips were developed by the Medipix collaboration at CERN \cite{f}. Detectors based on them, in addition to recording the coordinates of the particle interaction in a sensor, are capable of measuring the energy released during the interaction. The Timepix chip can operate in several modes:
\begin{itemize}
    \item Medipix - photon counting mode. In this mode, the chip counts how many times a signal exceeds a set threshold over a certain period of time.
    \item Time of Arrival (ToA) - the chip measures when a signal exceeds a set threshold.
    \item Time over threshold (ToT) - the chip measures signal duration over certain threshold.
\end{itemize}

\par Since the signal duration over the threshold is proportional to the collected charge and thus to the energy deposited by the photon, the ToT mode allows us to determine the energy deposit in the sensor~ \cite{g}.

\par CdTe with a thickness of 2 mm was chosen as the sensor material, because CdTe has a higher absorption coefficient compared to Si and GaAs. The main advantages and disadvantages of this material are perfectly described in the review of \cite{h}.

\par The detector area is 14.08~mm x 14.08~mm, which is significantly smaller compared to other gamma cameras. Therefore, the operation of the system for a small field of view must be ensured by a high-precision lensing system, which is provided by the coding aperture. In this work, we used a MURA-type coding aperture, with a matrix rank of 31 as a collimator. This type of collimator has a rectangular working area and, when rotated by 90 degrees, changes transparent elements to non-transparent ones and vice versa. This feature allows us to increase the signal-to-noise ratio \cite{j}. The collimator, 1~mm thick, was made of tungsten, and the radius of holes was 170~$\rm \mu$m.

\par The placement of the detector and the coding aperture is done by a duralumin mount, which allows adjusting the distance between them. The mount features a rotating platform, allowing quick rotation of the mask pattern by 90~degrees.

\par By changing the distance between the collimator and the detector, it is possible to change the field of view of the SPECT system:

\begin{equation}
    FoV = \frac{(f+d)(M-D)}{d} + D ,
\end{equation}
where $FoV$ - field of view size, $f$ - distance from the source to the collimator, $d$ - distance from the collimator to the detector, $M$ - the collimator size, $D$ - the detector size.

\par
\begin{figure}[!ht]
\centering 

\qquad
\includegraphics[scale = 0.5]{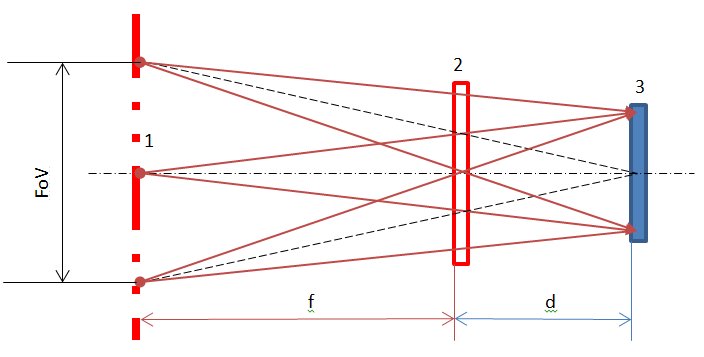}
\caption {\label {fig:1} Layout of the SPECT setup:
1~-~source plane, 2 - coding aperture, 3 - detector, {\it FoV} - field of view, {\it f}~-~distance from the source to the coding aperture, {\it d} - distance from the coding aperture to the detector}

\end{figure}

 In this work, the distance between the detector, the collimator, and the object under study was selected so that the field of view was 57.75~mm~x~57.75~mm~(Fig.~\ref{fig:1}).

\par The object was mounted on a dedicated rotating platform, allowing it to be turned around a single axis during imaging. This simplified the design of the experimental setup for phantom testing by eliminating the need for a rotating gantry. For SPECT characterization, 120~projections were acquired at different angular positions of the object.

\section{Data pre-processing}

\par Energy measurements in this study were conducted using the TOT mode of the Timepix chip. A pixel-by-pixel calibration was performed to convert the duration of signals over the threshold into a photon energy~\cite{g}, with a bias voltage of -450 V and an energy threshold of 6 keV.

\par Data from the Timepix detector were collected using the FitPix interface module \cite{Kraus}. The output data contained frames recording the coordinates of activated pixels and their energy deposition in TOT units. Neighboring activated pixels were grouped into clusters for further analysis.

\par Each aperture hole effectively operates as a pinhole collimator, creating a composite shadowgram that must be decoded to form the final image. The decoding step involves convolving the shadowgram with a decoding function \cite{j}. However, because of the detector's pixelated architecture and charge-sharing effects stemming from the 2-mm-thick CdTe sensor, data preprocessing is required. To address these challenges, specialized software was developed to process each frame, following these steps:

\begin{enumerate}
    \item Selection of events based on cluster size.
    \item Pixel-by-pixel energy calibration.
    \item Filtering of events based on photon energy.
    \item Reconstruction of photon interaction coordinates.
    \item Decoding of the shadowgram.
\end{enumerate}

\par Several factors determine the cluster size for photons with the energy up to 140 keV, including the diffusion of charge carriers during their drift to the readout electrodes, the fluorescent photons (whose free path in CdTe is up to 100 {\rm $\mu$}m \cite{Tlustos}, which is comparable in size with almost two Timepix pixels), and the Compton electrons. Additionally, at high fluxes or with a long acquisition time, signals from several photons can overlap, forming larger clusters. The bias voltage applied to the sensor, which governs the drift time of the charge carriers and the acquisition time, also indirectly affects the cluster size.

\par
\begin{figure}[!ht]
    \centering
    \includegraphics[width=0.7\linewidth]{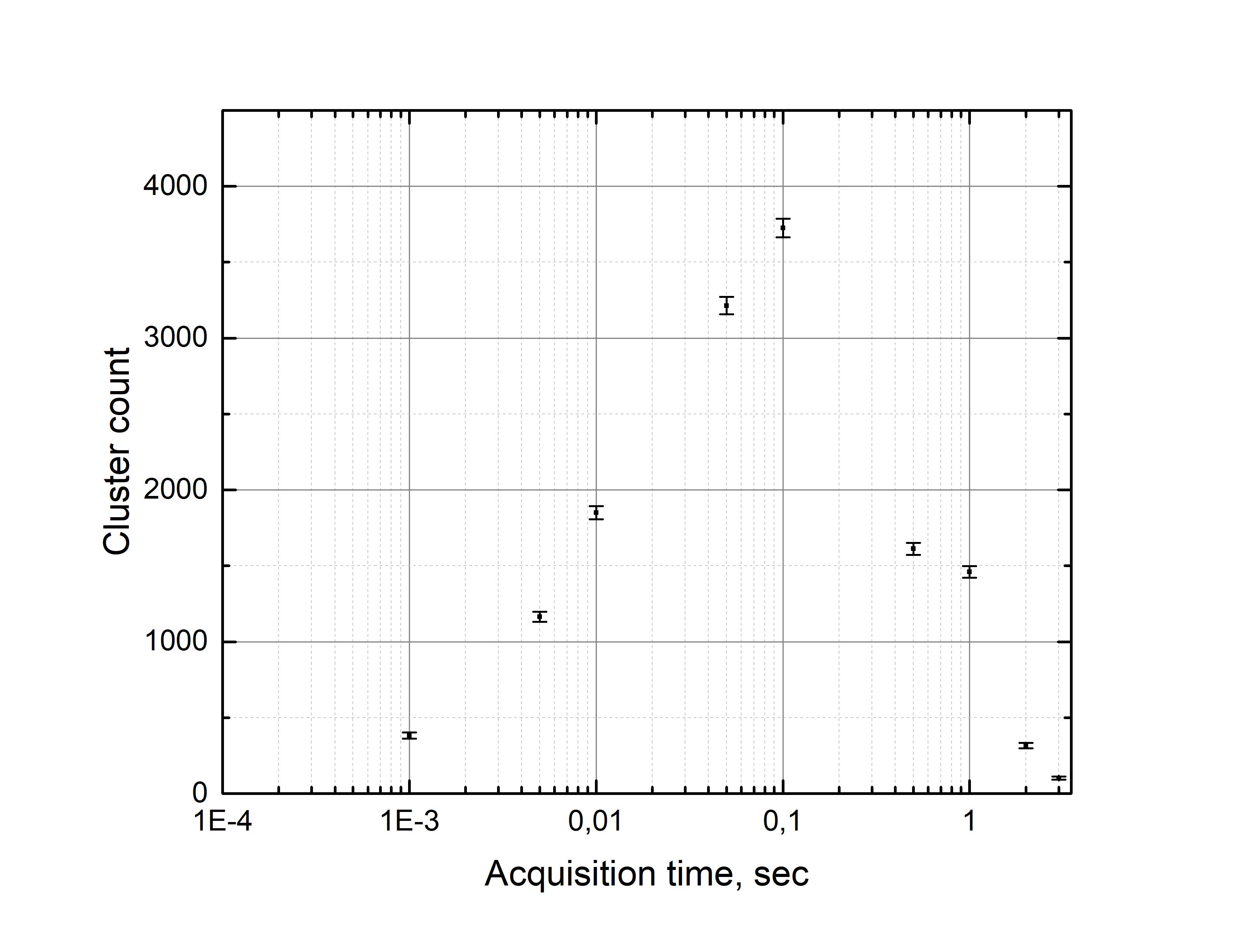}
    \caption{Dependence of the cluster count on the acquisition time.}
    \label{fig:aquisition_time}
\end{figure}

 Selecting an optimal acquisition time involves balancing the overlap of signals and the dead time due to frame readout. Overlapping signals can distort cluster formation and degrade projection quality, while shorter acquisitions, while reducing overlap, increase dead time, which is particularly undesirable with short-lived isotopes. Figure \ref{fig:aquisition_time} shows how the number of registered clusters depends on acquisition time per frame during a total measurement of one minute. Longer acquisition time leads to more registered events because of reduced dead time. However, when acquisition time goes above 0.1 seconds, the number of registered events decreases substantially due to increased overlaps forming larger clusters. Consequently, an acquisition time of 0.1 seconds was selected as optimal, yielding the highest number of events with minimal overlap.

\par
\begin{figure}[!ht]
    \centering
    \includegraphics[width=1\linewidth]{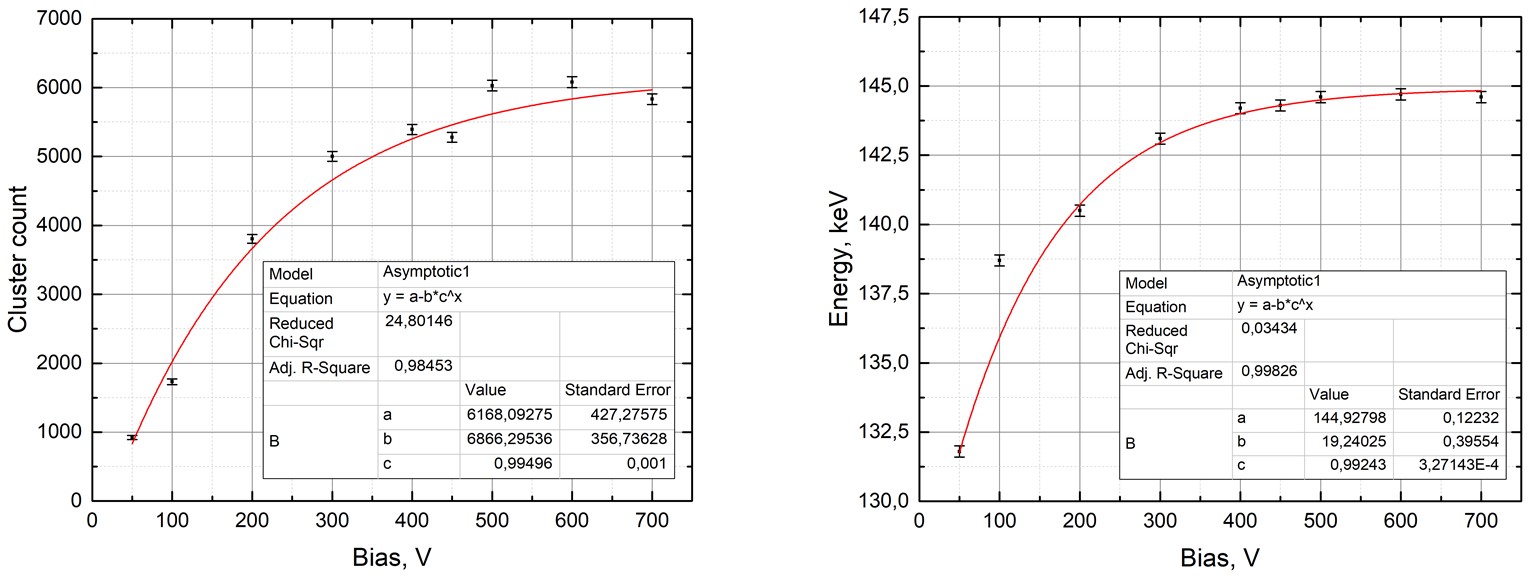}
    \caption{Dependency of the registered cluster count (left) and cluster energy (right), versus the bias voltage on the sensor.}
    \label{fig:bias}
\end{figure}

 The dependence of detection efficiency and reconstructed energy for photons from the decay of $^{99m}Tc$ (Fig. \ref{fig:bias}) indicates that the charge collection efficiency in a 2 mm thick CdTe sensor stabilizes at bias voltages above -400 V, approaching the asymptotic value by less than 7\%. Furthermore, the per-pixel energy calibration conducted at a bias voltage of -450 V remains applicable at higher voltages with an uncertainty less than 1\%. 

\par
\begin{figure}[!ht]
     \centering
     \includegraphics[width=0.7\linewidth]{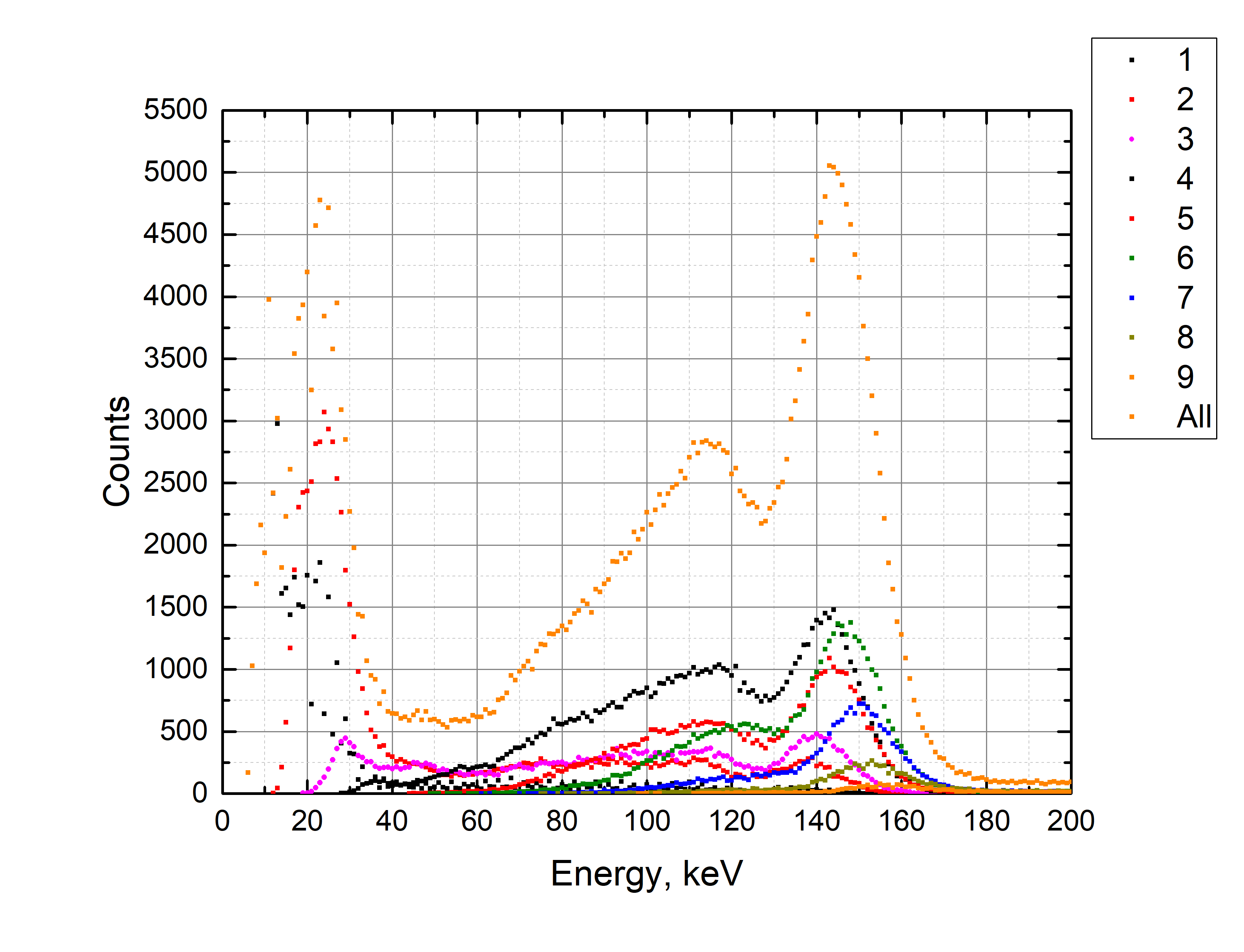}
     \caption{$^{99m}Tc$ energy spectra for various cluster size.}
     \label{fig:cluster_counts}
\end{figure}

 A small discrepancy is clearly visible between the measured energy (145 keV) and the photopeak energy of $^{99m}Tc$ (140.5 keV) in Fig. \ref{fig:cluster_counts}. This discrepancy is caused by the energy miscalibration for many-pixel clusters, since the existing calibration relies solely on single-pixel clusters. The maximum energy that was used for calibration was 70~keV. The induced charge on some pixels remains below the threshold and thus goes undetected, particularly at photon energy above 30 keV. Consequently, the reference energy used for calibration is defined by single-pixel clusters whose recorded charge is lower than the actual charge generated during the interaction of photons with the sensor. As the cluster size increases, the discrepancy also increases, since this small fraction of the charge is not visible.
 
\par Larger clusters contribute less to the $^{99m}Tc$ photopeak and more to the “tails” of the energy spectrum. These tails arise from the overlap of events in smaller clusters, encompassing clusters larger than nine pixels. Single-pixel and two-pixel clusters primarily represent fluorescent photons, but unlike two-pixel clusters, single-pixel clusters do not contribute to the $^{99m}Tc$ photopeak.

\par To identify $^{99m}Tc$ photons, clusters with the reconstructed energy within 22\% off the photopeak were selected, corresponding to the energy resolution of Timepix detector (at 140 keV). Clusters spanning 2–9 pixels can be 110–220 {\rm $\mu$}m in size, which substantially deteriorates the spatial resolution of the reconstructed images. To mitigate this issue, the cluster’s total charge is reassigned to the pixel with the highest recorded energy. Thus, before the image decoding, the data are converted into a two-dimensional array of single-pixel events with energies around the $^{99m}Tc$ photopeak region.

\section{System tests}

\par The methods of measuring main characteristics of the system, as contrast, sensitivity, uniformity, linearity, SNR, spatial resolution, energy resolution, etc., as well as the detector data preprocessing algorithm and clustering, and the methods for improving SNR using machine learning, have been published elsewhere~\cite{k,l,m,n}. This paper presents results on basic tests developed following the IAEA recommendations and relevant documents of national standard agencies~\cite{a, b, c, d}. The reconstructed projections were not subjected to any additional filtering, except for cluster preprocessing described above.

\subsection{Uniformity}

\par
\begin{figure}[!ht]
    \center{\includegraphics[scale=0.7]{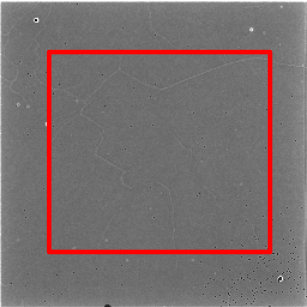}}
    \caption{Reconstructed image of the detector response to an X-ray source.}
    \label{fig:uniformity_xray}
\end{figure}

 Intrinsic uniformity provides a quantitative measure of potential artifacts within the field of view (FoV). To determine this parameter, a pointlike source is often used. In our case, an X-ray tube with a 75 {\rm $\mu$}m focal spot was used\footnote{SB 120-350 by Source Ray Inc.}, and the exposure time was set to 5 minutes. The analysis of the results is carried out in the central field of the detector, highlighted in red. This was done in order not to affect the edge sensors, such as leakage currents, which can significantly~(Fig. \ref{fig:uniformity_xray}).

\par
\begin{figure}[!ht]
\begin{minipage}[h]{0.4\linewidth}
\center{\includegraphics[scale=0.05]{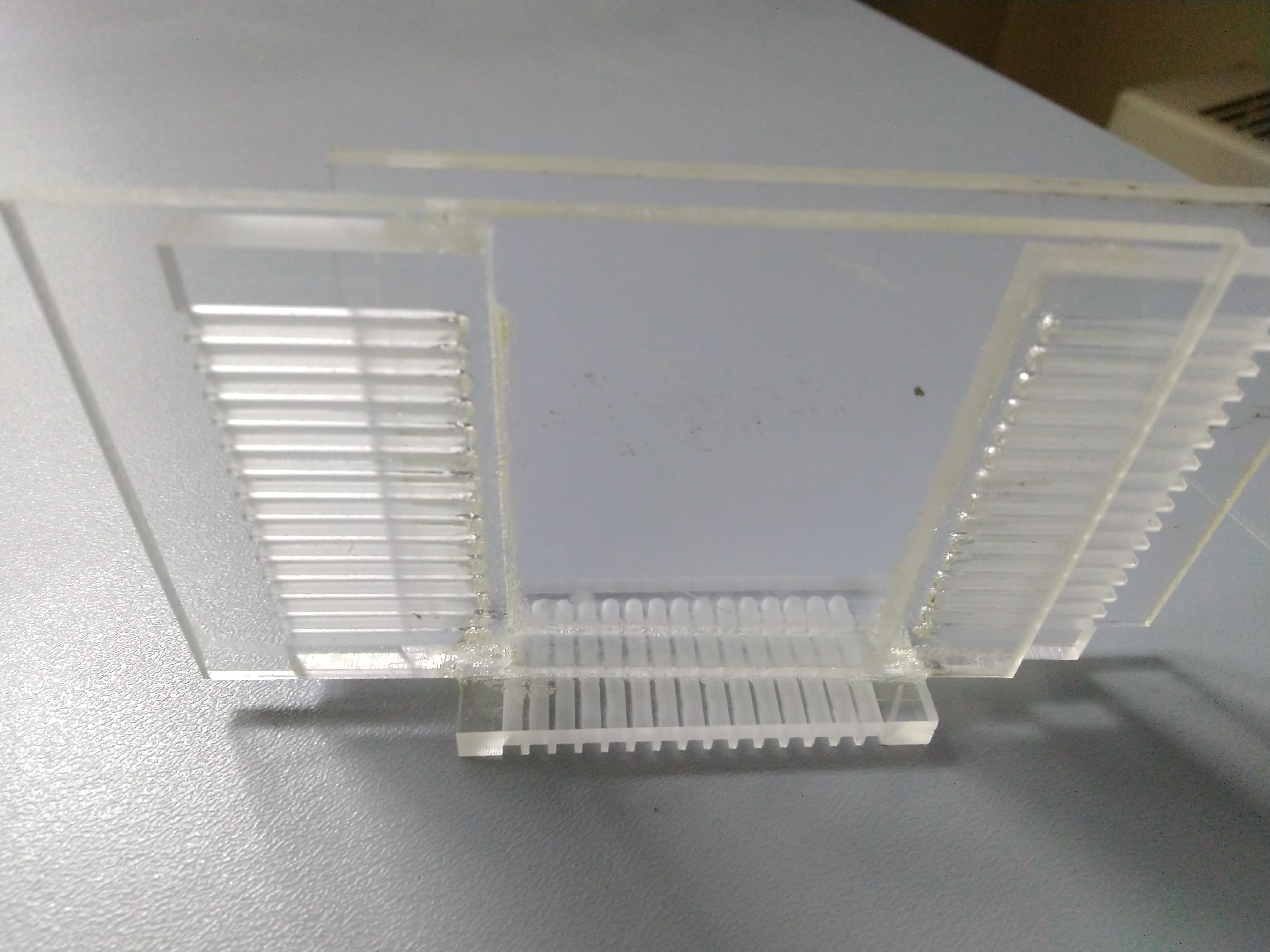}} \\
\end{minipage}
\hfill
\begin{minipage}[h]{0.4\linewidth}
\center{\includegraphics[scale=0.05]{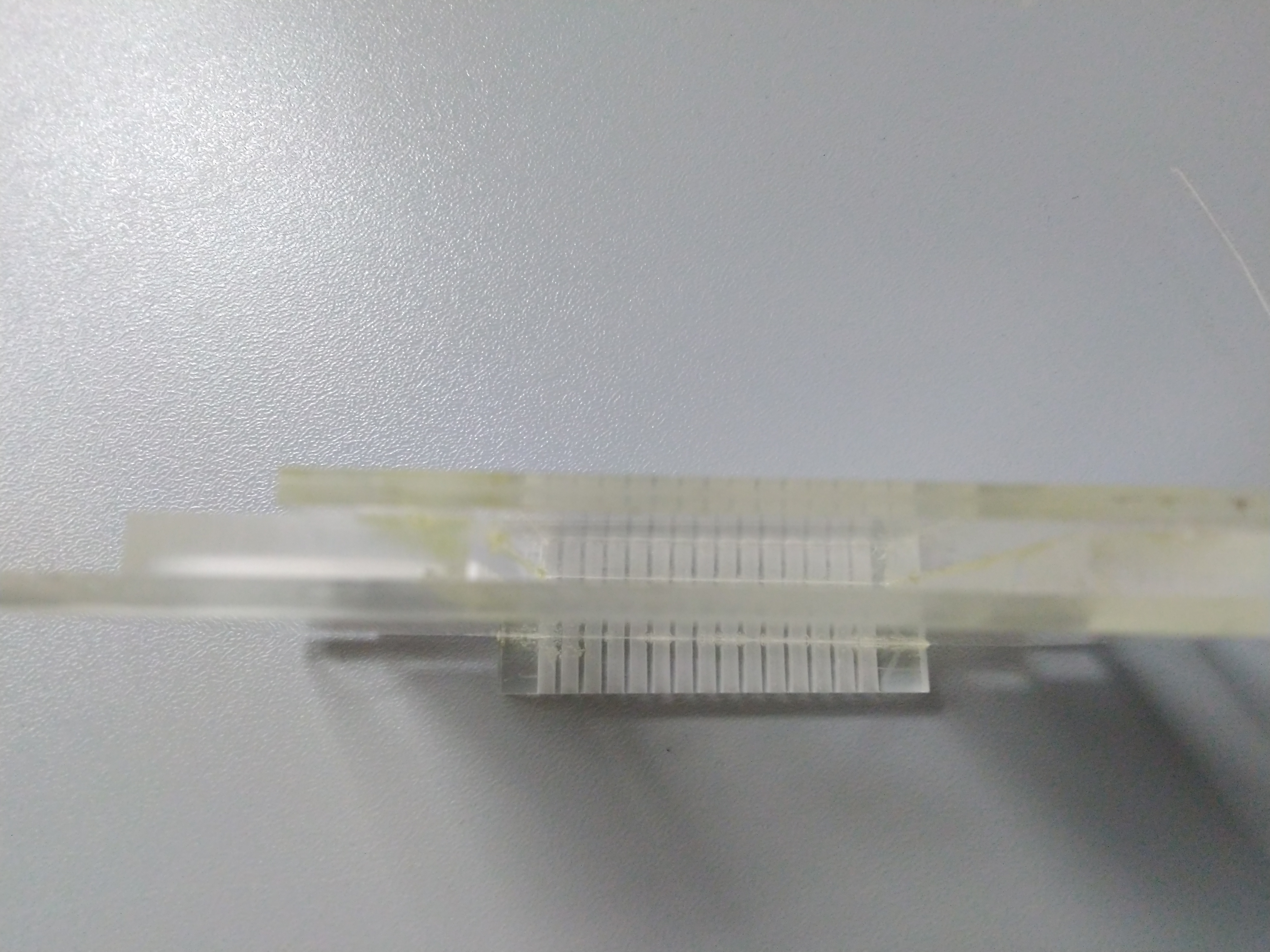}} \\
\end{minipage}
\caption{Flood phantom. Side view(left). Top view (right)}
\label{fig:Flood}
\end{figure}

\begin{figure}[!ht]
    \center{\includegraphics[scale=0.7]{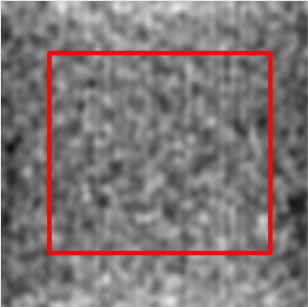}}
    \caption{Reconstructed image of a flood phantom.}
    \label{fig:uniformity_rec}
\end{figure}

 In addition to intrinsic uniformity, both integral and differential system uniformity were also determined. A flood phantom with the dimensions 45 $\times$ 50 $\times$ 5 mm$^3$ was used, positioned to cover as much of the 57 mm $\times$ 57 mm FoV as possible Fig. \ref{fig:Flood}. The detector regions used to determine intrinsic uniformity were selected for integral uniformity measurement, corresponding to the flood phantom’s placement. The~(Figure~\ref{fig:uniformity_rec}) shows an image of distributed sources from a flood phantom, the total activity of the $^{99m}$Tc solution in the phantom was 1 MBq. Given the low activity, a 2 $\times$ 2 binning was applied to the decoded image to increase the statistics in each pixel. 

\par Before assessing uniformity, any pixel whose count is below 75\% of the average number of pulses is set to zero. The non-zero pixels are then smoothed over nine points with the following weights:
\begin{center}
    \begin{tabular}{ccc}
        1 & 2 & 1 \\
        2 & 4 & 2 \\
        1 & 2 & 1 \\
    \end{tabular}
\end{center}

\par To calculate the integral uniformity, the following formula was used:
\begin{equation}
    \text{Integral uniformity} = \frac{\text{max} - \text{min}}{\text{max} + \text{min}},
    \label{eq:diff_uniformity}
\end{equation}
where $\text{max}$ is the maximum value of the total number of photons registered by the detector and $\text{min}$ is the minimum one. The integral uniformity was calculated only for the area covered by of the phantom image, neglecting narrow side bands of FoV.

\par Differential uniformity was also calculated using~\ref{eq:diff_uniformity}, but only for a small detector area (5 $\times$ 5 pixels), not for the entire pixel matrix. The maximum and minimum were evaluated for all possible vertical and horizontal groups.

\par Thus, the integral and differential uniformity was determined for the same sensor area. The intrinsic uniformity is 92\%. Integral uniformity and differential uniformity are 44\% and 84\%, respectively.

\par The low integral uniformity partially comes from the presence of artifacts caused by the shortcomings of the shadowgram reconstruction. The reconstruction algorithm currently does not take into account the circular shape of the collimator holes, assuming it to be square. The use of filters and the development of more advanced reconstruction methods can significantly reduce the number of artifacts and increase system integral uniformity.

\subsection{Linearity}

\par The linearity is the measure of geometry distortion of source distribution by the SPECT system in the object under study. Measuring linearity is especially important for SPECT systems equipped with scintillation detectors, because for them, the linearity will be affected by the inhomogeneity and defects of the crystal, as well as by the accuracy of the placement of PMTs. A linearity of 100 percent indicates that the reconstructed image does not distort the relative position of sources. 

\par
\begin{figure}[!ht]
\begin{minipage}[h]{0.28\linewidth}
\center{\includegraphics[scale=0.21]{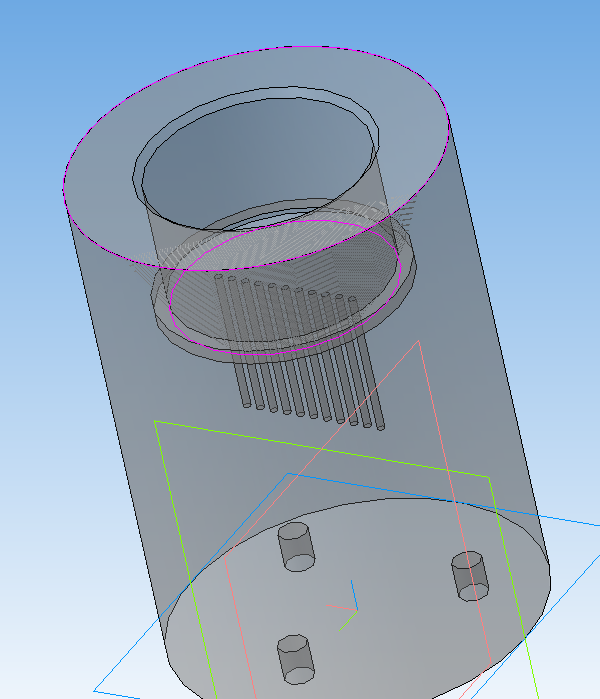}} \\a) \\
\end{minipage}
\begin{minipage}[h]{0.35\linewidth}
\center{\includegraphics[scale=0.29]{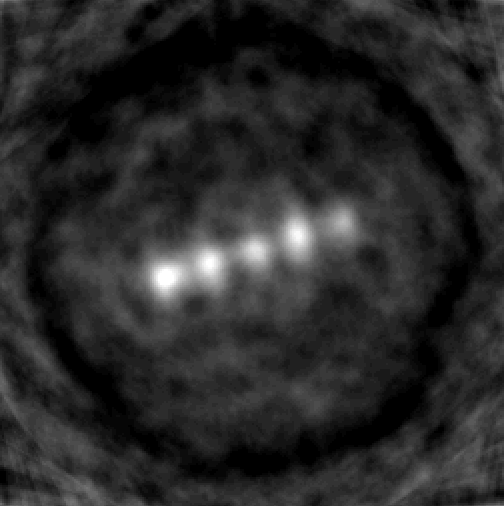}} \\b) \\
\end{minipage}
\begin{minipage}[h]{0.35\linewidth}
\center{\includegraphics[scale=0.27]{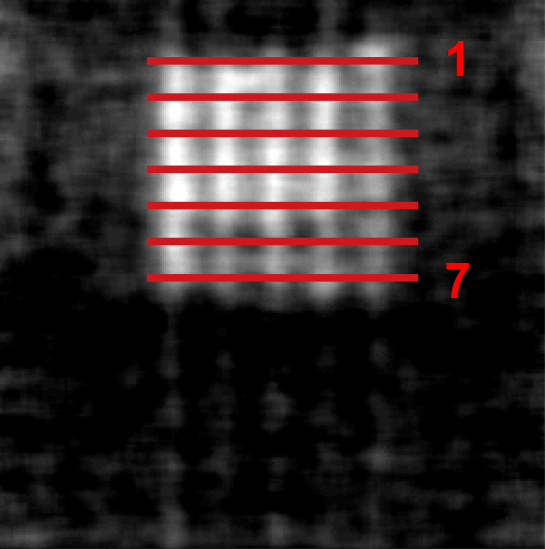}} \\c) \\
\end{minipage}
\caption{Linearity phantom: a) drawing, b) reconstructed slice, c) sagittal section. Lines correspond to measured profiles.}
\label{fig:2}
\end{figure}

 To determine the linearity, a phantom was used with capillaries aligned parallel in a row. The body of the phantom was made of plexiglass with a mechanical precision of 0.05 mm (Fig. \ref{fig:2}). Inside the phantom, there were 11 capillaries, each 1.1 mm in diameter. The distance between the centers of the capillaries was 2 mm, and each capillary was 20 mm long. The rotation axis of the phantom passed through the central capillary. All even capillaries were filled with $^{99m}Tc$ solution. The total activity was in 83~MBq. For the tomographic reconstruction, projection images were taken with an exposure time of 2 min per projection. 

\par
\begin{figure}[!ht]
\begin{minipage}[h]{0.45\linewidth}
\center{\includegraphics[scale=0.2]{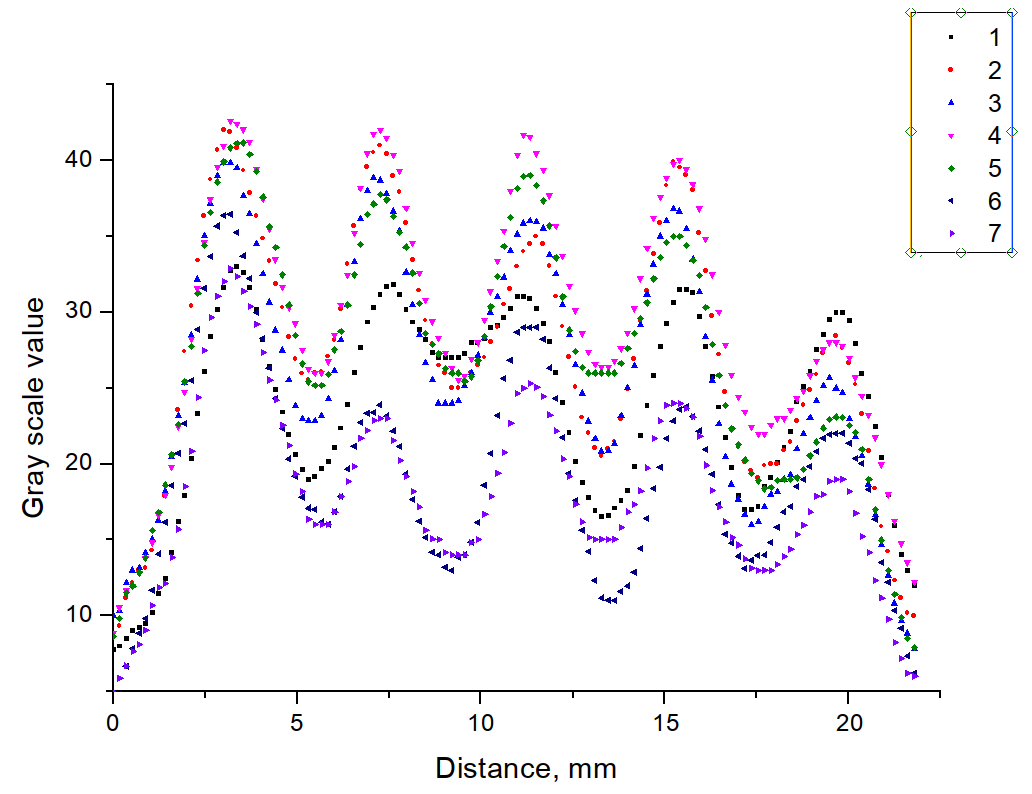}} \\a) \\
\end{minipage}
\begin{minipage}[h]{0.5\linewidth}
\center{\includegraphics[scale=0.25]{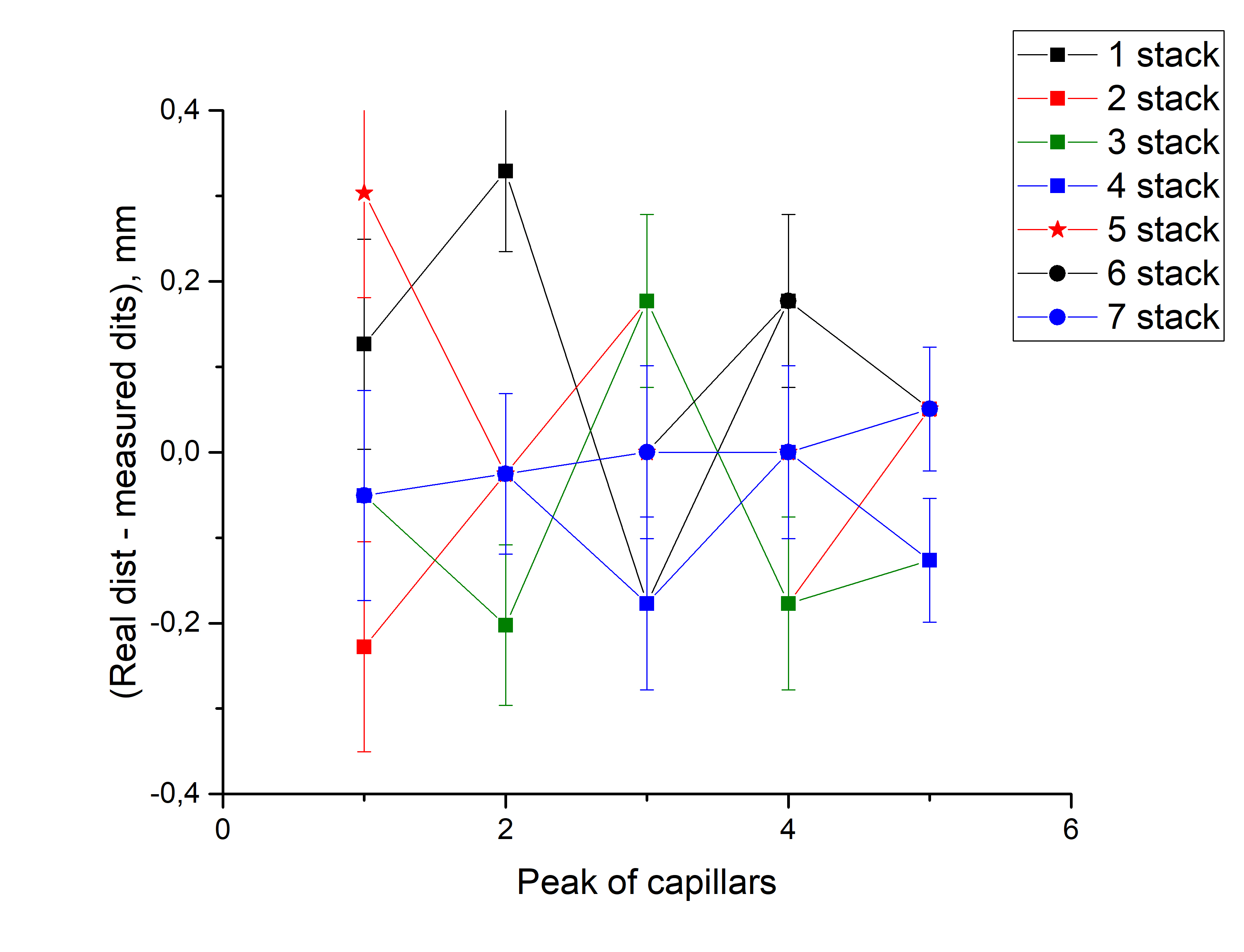}} \\b) \\
\end{minipage}
\caption{a) Capillary plot profiles measured by linearity phantom, b) peak position in section}
\label{fig:3}
\end{figure}

 After the reconstruction, a sagittal section was selected that passed through the centers of each of the capillaries (Fig. \ref{fig:3}). On this section, 7 profiles were constructed passing through each capillary along the entire height of the phantom. The location of the intensity maximum, corresponding to the center of each capillary was determined along each profile. Subsequently, these data were compared with the phantom's blueprint to determine the deviation of the measured center from its actual value. The deviations were less than 0.2 mm on a 20 mm length of the capillary, or, in other words, less than 1\%. Such high accuracy is ensured by the high precision of the collimator pattern manufacture, 0.001 mm, as well as the high precision of the pixel matrix placement of the detector.

\par
\begin{figure}[!ht]
\begin{minipage}[h]{0.33\linewidth}
\center{\includegraphics[scale=0.25]{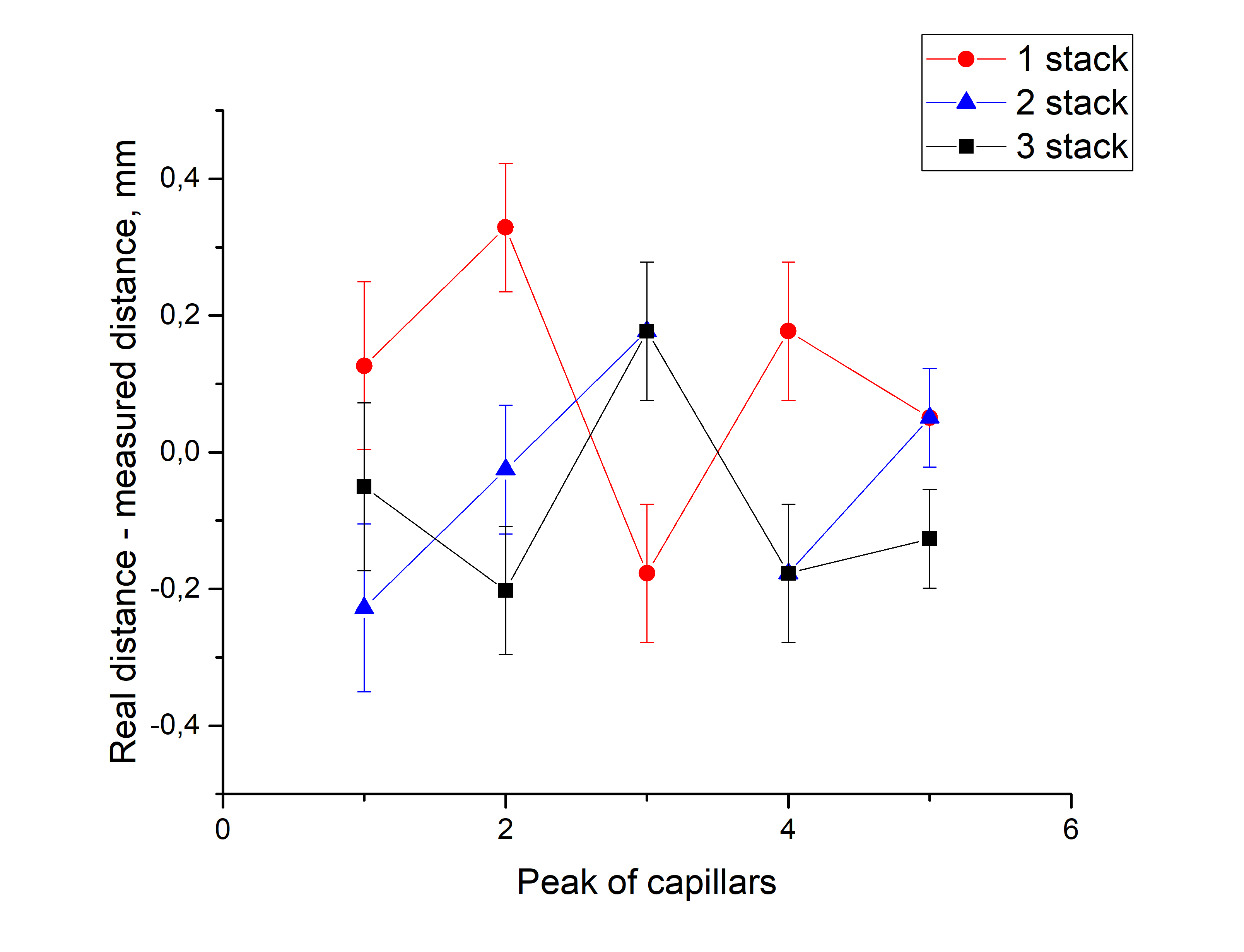}} \\a) \\
\end{minipage}
\begin{minipage}[h]{0.6\linewidth}
\center{\includegraphics[scale=0.25]{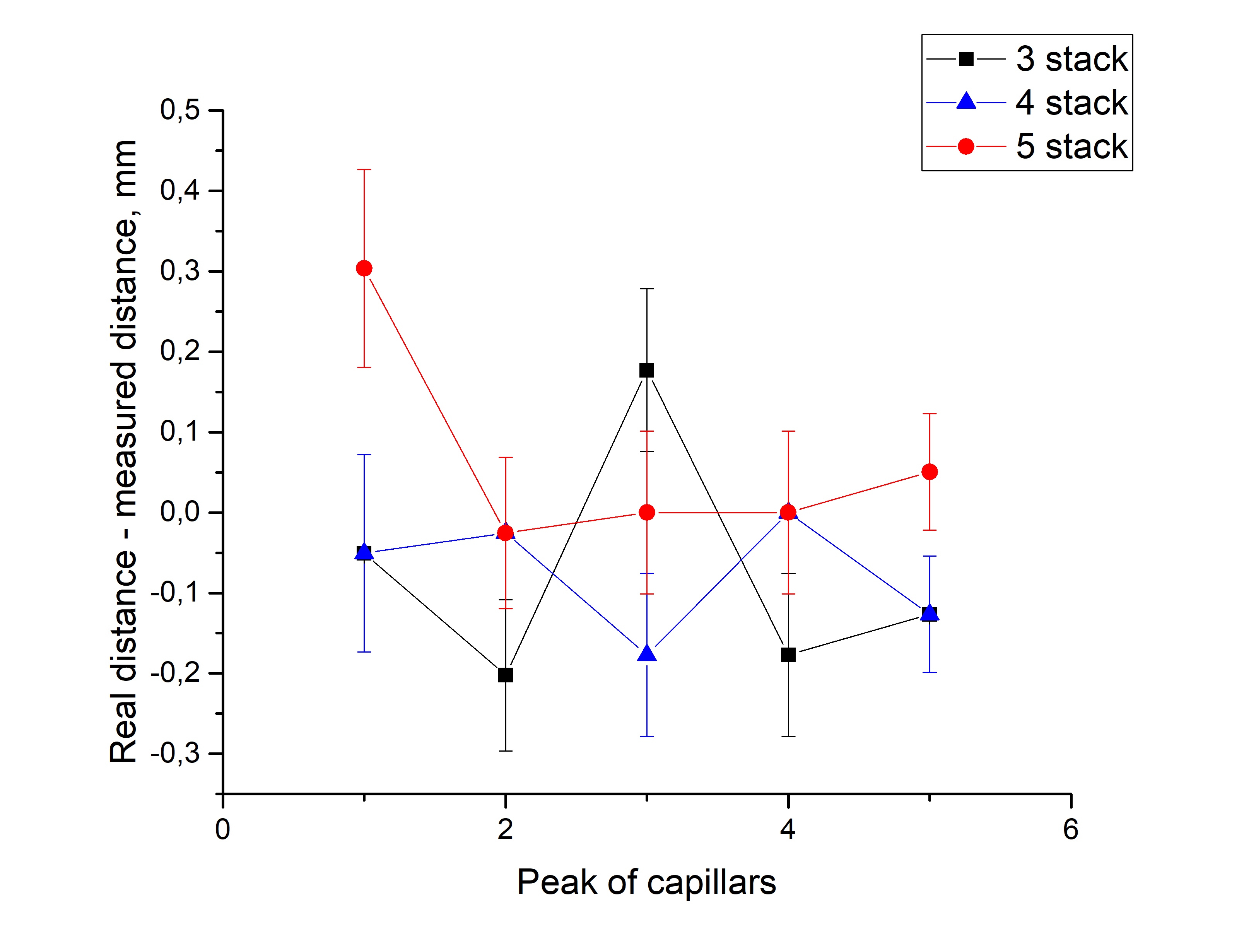}} \\b) \\
\end{minipage}
\begin{minipage}[h]{0.9\linewidth}
\center{\includegraphics[scale=0.25]{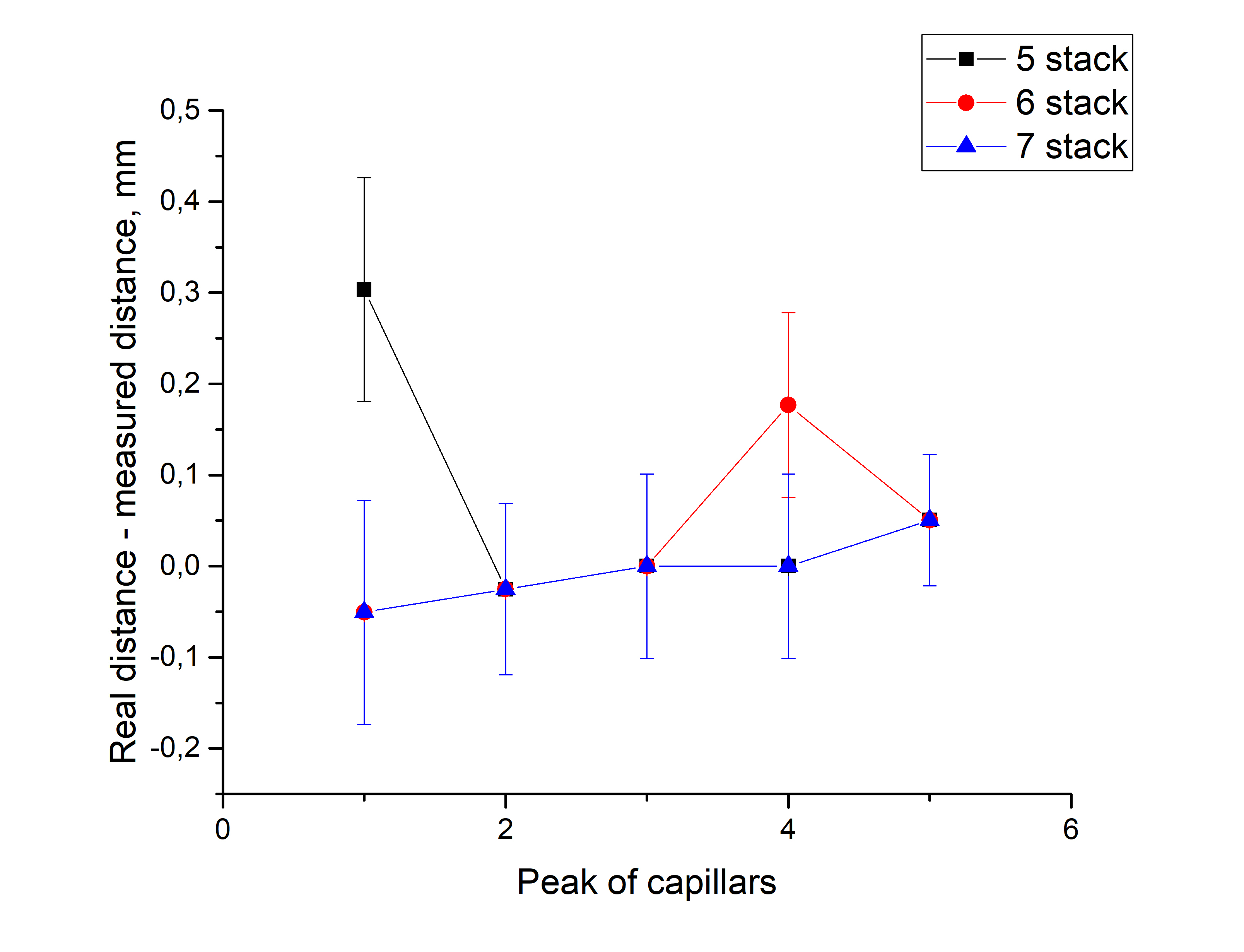}} \\c) \\
\end{minipage}
\caption{Linearity phantom profiles: a) 1-3 slices, b) 3-5 slices, c) 5-7 slices.}
\label{fig:linearity_slices}
\end{figure}

 The data for the triplets of linearity phantom slices are presented in Fig.~\ref{fig:linearity_slices}.

\subsection{Spatial resolution}

\par Unlike the initial system tests conducted with a field of view of 30 mm x 30 mm for planar images~\cite{k}, this series of experiments aimed at the study of tomographic spatial resolution and involved 3D reconstruction of distributed sources for a field of view of 57 mm $\times$ 57 mm. Due to the pixel structure of the sensor, the increase in the field of view causes the spatial resolution to decrease proportionally. 

\par
\begin{figure}[!ht]
\begin{minipage}[h]{0.95\linewidth}
\center{\includegraphics[scale=0.4]{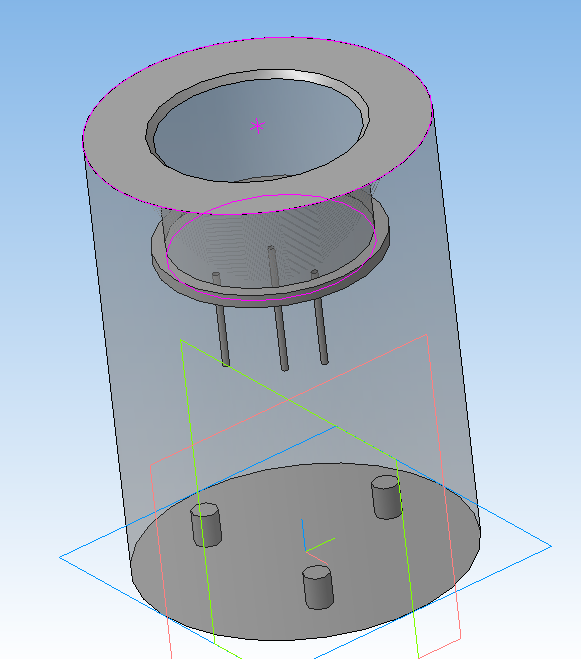}} 
\end{minipage}

\caption{Scheme of the phantom for measuring spatial resolution}
\label{fig:SR_phantom}
\end{figure}

 The following phantom was used to measure spatial resolution. The central capillary's axis passed through the phantom's rotation axis. At an equal distance from the central capillary, two other capillaries were aligned parallel, so that their centers formed an isosceles right triangle. The diameter of each capillary was 1.1 mm (Fig. \ref{fig:SR_phantom}).

\par
\begin{figure}[!ht]
\begin{minipage}[h]{0.5\linewidth}
\center{\includegraphics[scale=0.53]{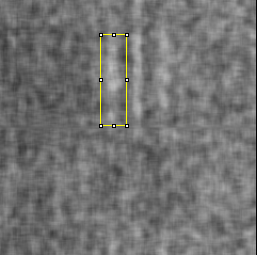}} \\
\end{minipage}
\hfill
\begin{minipage}[h]{0.5\linewidth}
\center{\includegraphics[scale=0.47]{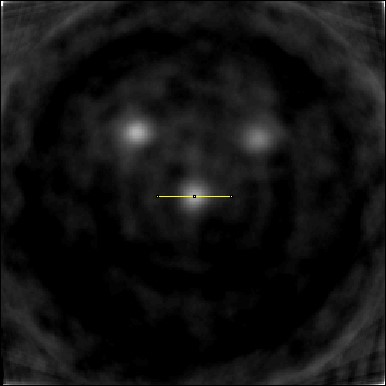}} \\
\end{minipage}
\caption{Original projections~(left) and slice~(right) of the phantom data}
\label{fig:proj_and_slice}
\end{figure}

 From figure ~\ref{fig:proj_and_slice}, it can be seen that capillaries with sources are hardly distinguishable on the projection. However, even such projections allow satisfactory tomographic reconstruction of distributed sources.

\begin{figure}[!ht]
\begin{minipage}[h]{0.5\linewidth}
\center{\includegraphics[scale=0.53]{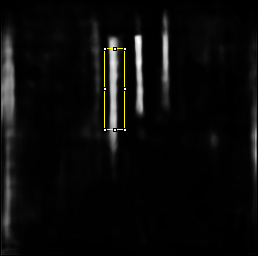}} \\
\end{minipage}
\hfill
\begin{minipage}[h]{0.5\linewidth}
\center{\includegraphics[scale=0.47]{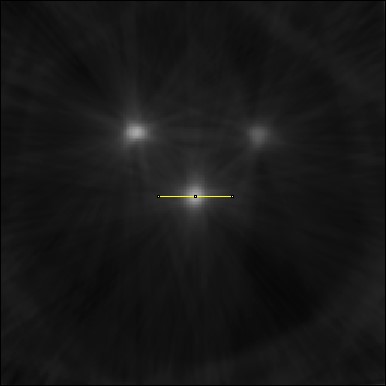}} \\
\end{minipage}
\caption{Projections~(left) and slices~(right) of the phantom data filtered using CED algorithm}
\label{fig:proj_and_slice_CED}
\end{figure} 

\begin{figure}[!ht]
\begin{minipage}[h]{0.48\linewidth}
\center{\includegraphics[scale=0.3]{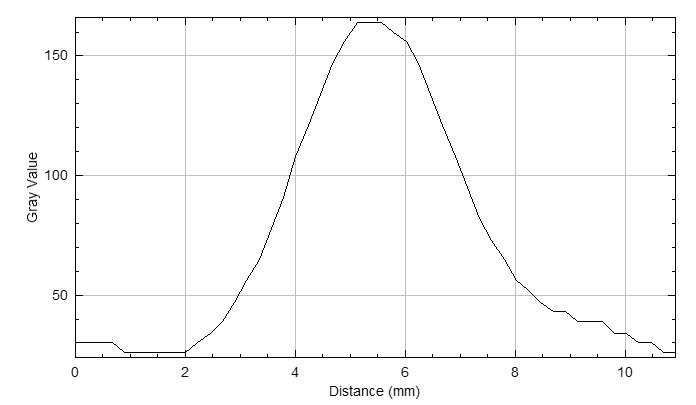}} \\a) \\
\end{minipage}
\begin{minipage}[h]{0.48\linewidth}
\center{\includegraphics[scale=0.29]{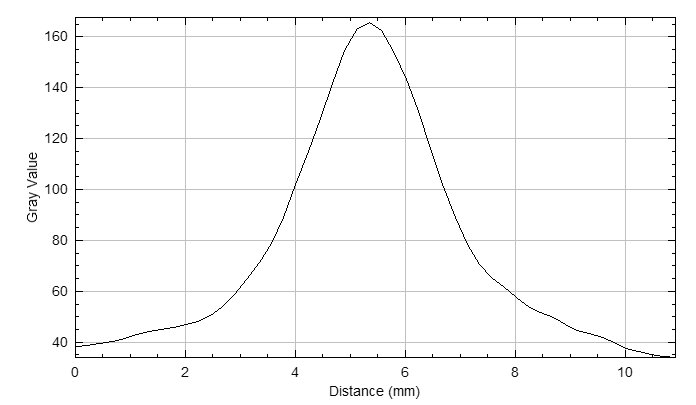}} \\b) \\
\end{minipage}

\caption{Original capillary plot profile (left) and capillary plot profile after CED algorithm (right).}
\label{fig:plot_profiles}
\end{figure}

\begin{figure}[!ht]
\begin{minipage}[h]{0.48\linewidth}
\center{\includegraphics[scale=0.3]{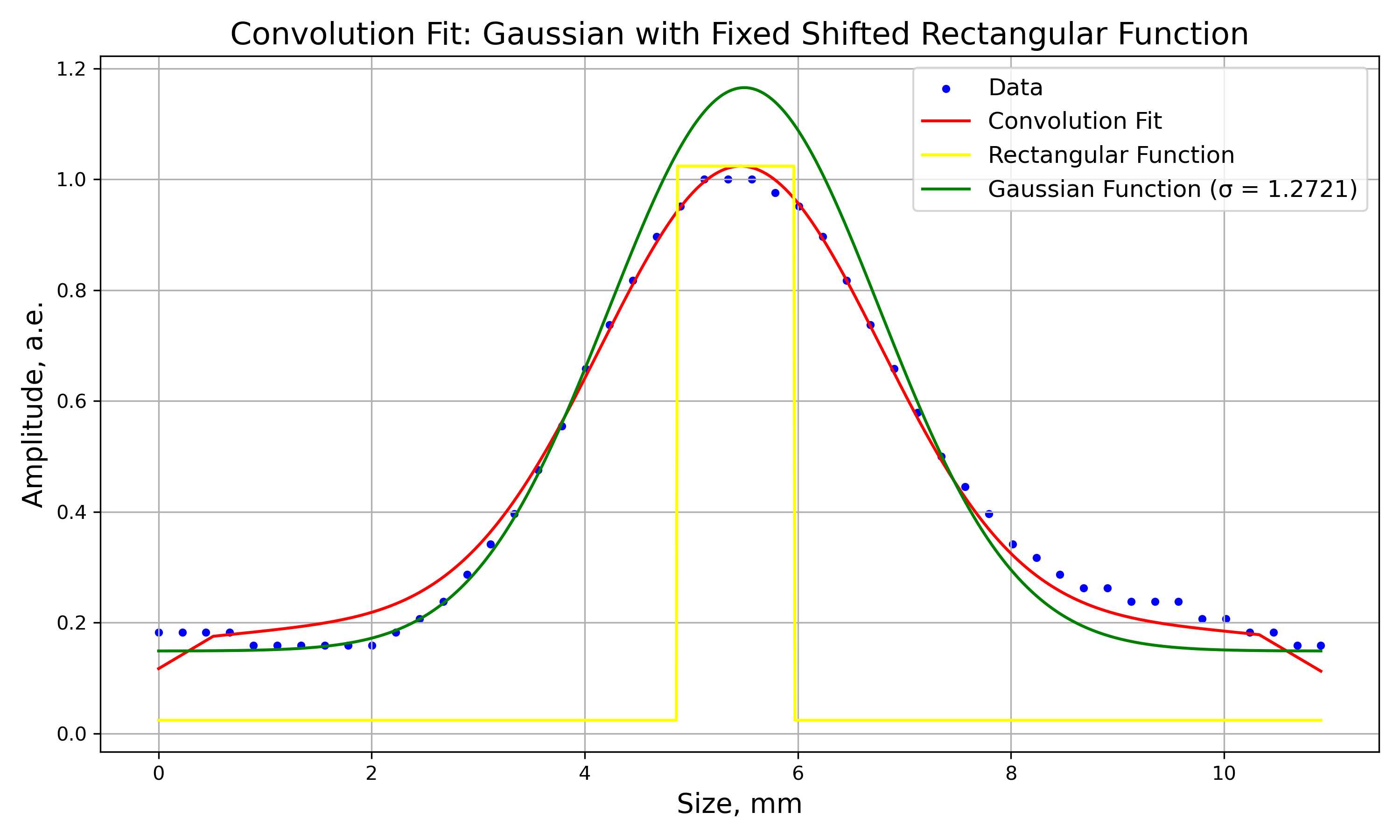}} \\a) \\
\end{minipage}
\begin{minipage}[h]{0.48\linewidth}
\center{\includegraphics[scale=0.29]{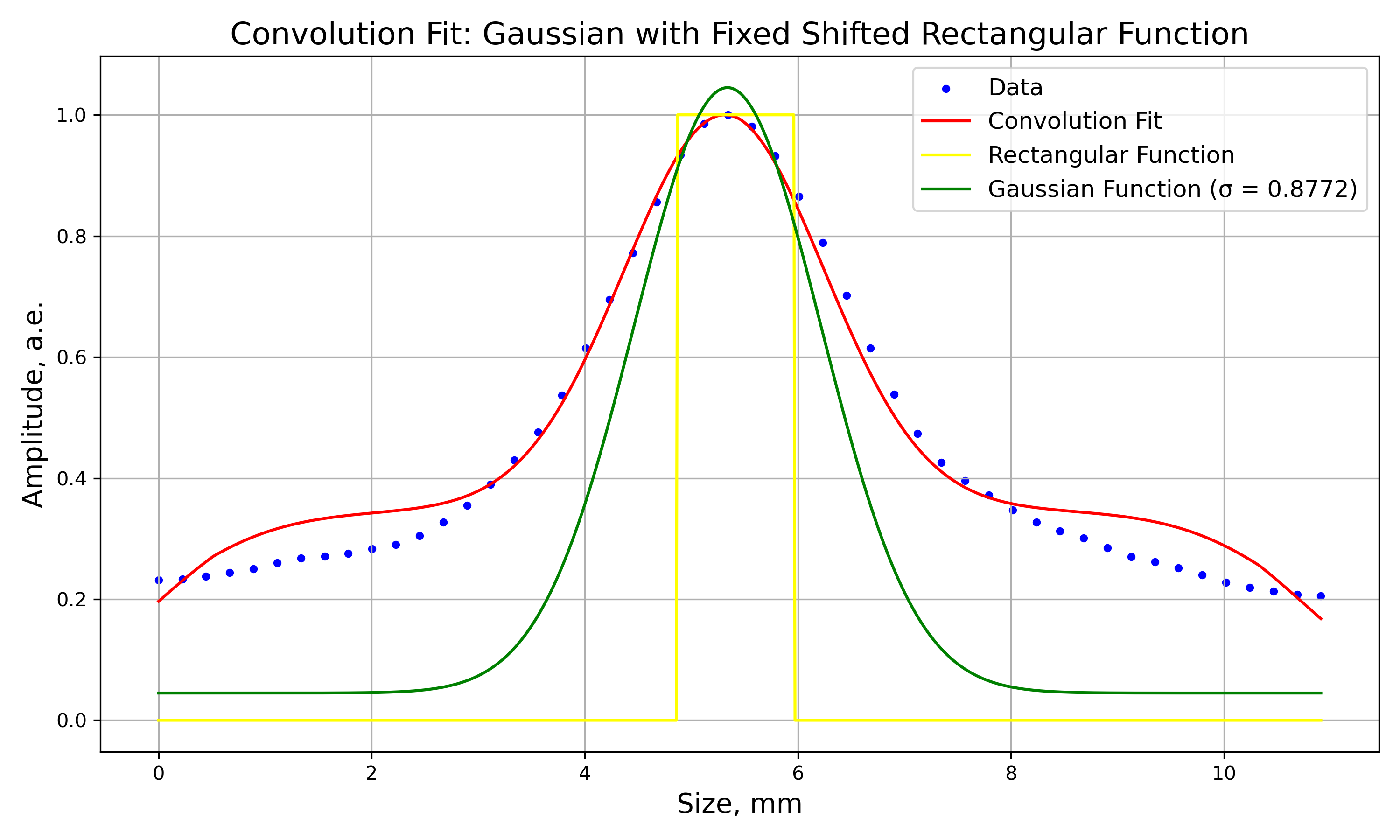}} \\b) \\
\end{minipage}

\caption{Capillary profile without filtering(left) and after CED-algorithm filtration (right).}
\label{fig:plot_profiles_CED}
\end{figure}

\par The degradation of the SNR is caused by sources located behind and in front of the focal plane~\cite{p}. To improve the quality of projection data, filtration can be applied or, as shown in the publication \cite{o}, machine learning methods can be used to improve significantly the SNR. For projections filtered by the CED-line algorithm, the FWHM of the capillary profile is shown to be consistent with the size of the capillary~(Fig.~\ref{fig:proj_and_slice_CED}).

\par To determine the tomographic spatial resolution (the Full Width at Half Maximum, FWHM), the capillary profile in the reconstructed slice was fitted by the convolution of a rectangular function and a Gaussian function. The width of the rectangular function was fixed to be 1.1 mm, which is the diameter of the capillary. The parameter $\sigma$ of the Gaussian function was free and fitted to determine the resolution. The result is shown in figure~\ref{fig:plot_profiles} for original profile and after applying the CED algorithm. The spatial resolution is 3.0 mm for the original profile, and improves up to 2.0 mm after filtration figure~\ref{fig:plot_profiles_CED}, which is consistent with our previous results obtained in 2D geometry. 
 
\subsection{Contrast}

\par
\begin{figure}[!ht]
    \center{\includegraphics[scale=0.4]{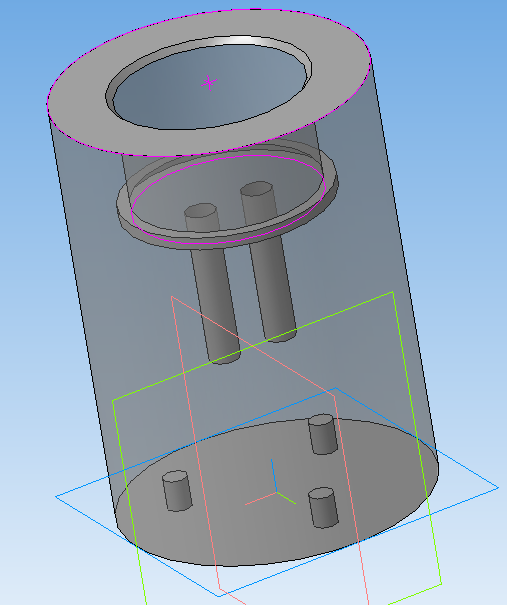}}
    \caption{Phantom used for contrast evaluation.}
    \label{fig:phantom}
\end{figure}

\begin{figure}[!ht]
\begin{minipage}[h]{0.48\linewidth}
\center{\includegraphics[scale=0.4]{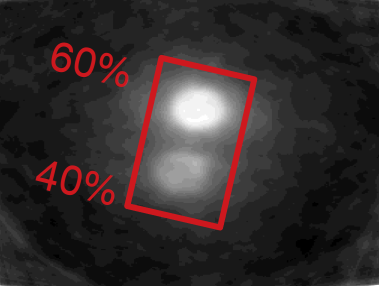}} \\a) \\
\end{minipage}
\begin{minipage}[h]{0.48\linewidth}
\center{\includegraphics[scale=0.29]{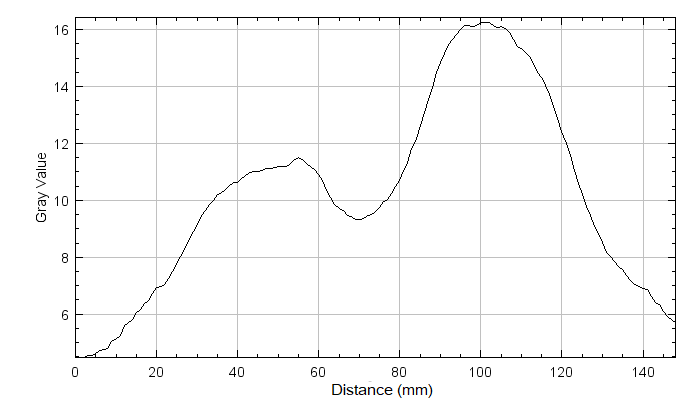}} \\b) \\
\end{minipage}

\caption{Contrast phantom slice~(left). Contrast phantom profile~(right)}
\label{fig:CP}
\end{figure}

 To evaluate the contrast of the SPECT system, a phantom consisting of two identical cells equidistant from the center of rotation was used (Fig.~\ref{fig:phantom}). The distance between cell centers was 10 mm. The total activity introduced into the phantom was 75 MBq. The cells were filled up with the solutions of 40\% and 60\% of the total activity, respectively. After reconstruction, the stack of slices was summed, and the intensity for the two cells were determined from the resulting profile. The ratio of the cell intensity was equal 0.67, closely matching the activity placed into the cells (Fig. \ref{fig:CP}). Thus, the contrast of the system is sufficiently high and accurate, making the system suitable for quantitative imaging.

\subsection{Sensitivity}
\par Sensitivity is defined as the ratio of the number of photons registered by the system to the number of photons emitted within a solid angle covered by the detector. The sensitivity can be estimated for a sample without a collimator (intrinsic sensitivity) and with the collimator (system sensitivity).

\par The 2 mm diameter capillary-shaped $^{99m}$Tc source with an activity of 100 MBq was located at a distance of 215 mm from the collimator. System sensitivity can be estimated using the formula:
\begin{equation}
   {S = \varepsilon_{\text{geom}} \cdot \varepsilon_{\text{att}} \cdot \varepsilon_{\text{det}} \cdot f_{\text{col}} \cdot A},
    \label{eq:sensitivity}
\end{equation}

\noindent where: $S$ — system sensitivity, [cps/MBq], $\varepsilon_{\text{geom}}$~—~geometric efficiency (the solid angle covered by the detector), $\varepsilon_{\text{att}}$~—~attenuation factor in air~\ref{eq:atten}, $\varepsilon_{\text{det}}$~—~detection efficiency (including absorption in the crystal and other losses), $f_{\text{col}}$~—~collimator efficiency (fraction of photons passing through collimator holes without scattering), $A$~—~source activity, [MBq].

\begin{equation}
    \varepsilon_{\text{att}} = e^{-\mu_{\text{air}} \cdot x}
    \label{eq:atten}
\end{equation}
\noindent where $\mu_{\text{air}}$ is the linear attenuation coefficient of air, and $x$ is the distance to the source,

\par The decrease of the activity caused by the source decay is not taken into account in the calculations since the acquisition time~(10 sec) is significantly less than the half-life of $^{99m}$Tc~(362~min).

\par The system sensitivity calculated by formula \eqref{eq:sensitivity} is equal to 79~cps/MBq. However, the total efficiency in 4$\pi$ solid angle, largely determined by a small detector surface (the solid angle covered by the detector is 0.0043 sr), is only 0.034\%. Also, 0.25\% of photons are lost due to absorption in the air. The factors influencing the system efficiency include the detection efficiency of the detector and the collimator geometry. The collimator geometry allows passing through as much as 39\% of incoming particles. In turn, the detection efficiency of photons with an energy of 140 keV is 60\% for a 2 mm thick CdTe sensor.

\section{Conclusion}

\par The results of this study demonstrate the feasibility of creating a compact SPECT system based on the Timepix detector with a 2~mm thick CdTe sensor and with a MURA-type coded aperture. The tests and experiments conducted included assessing the system's uniformity, linearity, spatial resolution, and contrast, as well as comparing performance under various filtering and reconstruction approaches.

\begin{enumerate}
    \item We evaluated intrinsic, integral, and differential uniformity, equal to 92\%, 44\% and 84\%, respectively. The integral uniformity metrics can be further improved by using better shadowgram reconstruction algorithm.
    \item Measurements of linearity using a capillary phantom confirmed the absence of significant distortions in the reconstructed images, indicating high accuracy in coordinate determination.
    \item Spatial resolution (FWHM) was determined to be 3.0 mm for the FoV of 57 mm $\times$ 57 mm. More advanced data processing and decoding demonstrated the potential to achieve a spatial resolution of about 2.0 mm.
    \item Experiments with a "hot" phantom (different amount of the radiotracer in equal volumes) confirmed that the system can distinguish areas with varying activity levels, closely matching the theoretical ratio. Which indicates high contrast of the system.
    \item The system’s overall sensitivity is primarily limited by the small detector area and certain design features of the collimator.
\end{enumerate}

\par The results obtained confirm the suitability of the prototype SPECT system for conducting preclinical experiments in a small field of view using $^{99m}Tc$-based radiopharmaceuticals. Possible improvements in reconstruction methods, the use of machine learning algorithms for data processing, and further optimization of collimator and transducer geometry can significantly improve image quality.
 






\end{document}